\documentclass[notitlepage,twocolumn,superscriptaddress,longbibliography,aps,nofootinbib]{revtex4-1}

\makeatletter
\def\@bibdataout@aps{%
\immediate\write\@bibdataout{%
@CONTROL{%
apsrev41Control%
\longbibliography@sw{%
    ,author="08",editor="1",pages="1",title="0",year="1"%
    }{%
    ,author="08",editor="1",pages="1",title="",year="1"%
    }%
  }%
}%
\if@filesw \immediate \write \@auxout {\string \citation {apsrev41Control}}\fi 
}
\makeatother

\usepackage[utf8]{inputenc}
\usepackage{amsmath,amsthm,amssymb,amsfonts,mathtools}
\usepackage{graphicx}
\usepackage[breaklinks=true,colorlinks=true,linkcolor=blue,urlcolor=blue,citecolor=blue,pagebackref=false]{hyperref}
\usepackage{color}
\usepackage{times}

\usepackage{notes2bib}
\bibnotesetup{
note-name = ,
use-sort-key = false
}

\usepackage{multirow,hhline}
\usepackage[table]{xcolor}

\usepackage{float}

\usepackage[normalem]{ulem}

\renewcommand{\vec}[1]{\mathbf{#1}}

\newcommand{\ie}{{\it i.e.},\ }
\newcommand{\eg}{{\it e.g.},\ }

\newcommand{\Htot}{\hat H_{\mathrm{tot}}}
\newcommand{\Ha}{\hat H_\mathrm{A}}
\newcommand{\Hf}{\hat H_\mathrm{f}}
\newcommand{\Hi}{\hat H_\mathrm{int}}

\newcommand{\pf}{f^\mathrm{s}} %

\newcommand{\ket}[1]{\left| {#1} \right\rangle}
\newcommand{\bra}[1]{\left\langle {#1} \right|}

\newcommand{\ketbra}[2]{\left| {#1}\vphantom{#2} \right\rangle\!\left\langle {#2}\vphantom{#1} \right|}

\newcommand{\ii}{\mathrm{i}}
\newcommand{\ee}[1]{{e}^{#1}} %

\newcommand{\RR}{\mathbb{R}}

\newcommand{\eup}{\mathrm{e}} %
\newcommand{\gup}{\mathrm{g}} %

\newcommand{\integral}[3]{\int_{#2}^{#3} \!\! \mathrm{d} #1 \,}

\newcommand{\spl}{\sigma^+}
\newcommand{\smi}{\sigma^-}
\newcommand{\w}{\omega}
\newcommand{\adop}{a^\dagger}
\newcommand{\aop}{a}
\newcommand{\hc}{\mathrm{h.c.}}
\newcommand{\im}{\mathrm{i}}

\newcommand{\be}{\begin{equation}}
\newcommand{\bea}{\begin{eqnarray}}
\newcommand{\eea}{\end{eqnarray}}
\newcommand{\pare}[1]{\left(#1\right)}
\newcommand{\spare}[1]{\left [#1\right ]}

\newcommand{\ts}{\tau_\mathrm{s}}

\begin{document}

\title{Non-perturbative treatment of giant atoms using chain transformations}
\author{David D. Noachtar}
\email{david.noachtar@tum.de}
\affiliation{Max-Planck-Institut für Quantenoptik, Hans-Kopfermann-Str.~1, 85748 Garching, Germany}
\affiliation{Technische Universität München, Am Coulombwall~3, 85748 Garching, Germany}

\author{Johannes Knörzer}
\email{johannes.knoerzer@eth-its.ethz.ch}
\affiliation{Max-Planck-Institut für Quantenoptik, Hans-Kopfermann-Str.~1, 85748 Garching, Germany}
\affiliation{Munich Center for Quantum Science and Technology, Schellingstr.~4, 80799 München, Germany}

\author{Robert H. Jonsson}
\email{robert.jonsson@mpq.mpg.de}
\affiliation{Max-Planck-Institut für Quantenoptik, Hans-Kopfermann-Str.~1, 85748 Garching, Germany}

\begin{abstract}
Superconducting circuits coupled to acoustic waveguides have extended the range of phenomena that can be experimentally studied using tools from quantum optics.
In particular giant artificial atoms permit the investigation of systems in which the electric dipole approximation breaks down and pronounced non-Markovian effects become important.
While previous studies of giant atoms focused on the realm of the rotating-wave approximation, we go beyond this and perform a numerically exact analysis of giant atoms strongly coupled to their environment, in regimes where counterrotating terms cannot be neglected.
To achieve this,
we use a Lanczos transformation to cast the field Hamiltonian into the form of a one-dimensional chain and employ matrix-product state simulations.
This approach yields access to a wide range of system-bath observables  and to relatively unexplored parameter regimes.
\end{abstract}

\maketitle

\section{Introduction}
Quantum optical theory provides a solid framework for the study of light-matter interaction.
Yet paradigmatic models are based on several approximations, such as the rotating-wave, electric dipole and Born-Markov approximations~\cite{Tannoudji2010}.
While the underlying assumptions are typically well justified, recent experimental advances have paved the way for investigations of yet unexplored parameter and physical regimes.
Superconducting circuits offer a versatile platform for such studies in which artificial atoms may be efficiently and strongly coupled to electromagnetic and sound waves~\cite{Kjaergaard2020}.
In particular \textit{giant atoms} challenge standard approximations and can only be accurately described when taking the finite spatial extent of the artificial atom into account \cite{ga_review_kockum}, plus a finite propagation speed if coupled to sound waves~\cite{Aref2016,Delsing2019} and counter-rotating terms beyond the rotating-wave approximation (RWA) at strong couplings.
Recent work has already capitalized on this and demonstrated several intruiging effects that occur in giant atomic setups, including decoherence-free interactions~\cite{Kockum2018}, non-exponential atomic decay~\cite{Andersson2019}, oscillating bound states~\cite{Guo2020} and chiral atom-waveguide couplings~\cite{Soro2021}.
Still theoretical treatment has so far been restricted to  couplings in the realm of the RWA.

At elevated light-matter couplings several physical phenomena can only be accurately captured by taking multiple field modes into account~\cite{Sundaresan2015,George2016,Gely2017}.
In this regime unphysical properties of single-mode models become more apparent such as, \eg causality violations in the form of superluminal signaling~\cite{zoharFermiProblemDiscrete2011,benincasaQuantumInformationProcessing2014,Jonsson2014,Munoz2018}.
In contrast to the single-mode quantum Rabi model (QRM)~\cite{Braak2011}, the corresponding multimode problem is not known to be integrable and requires novel techniques for theoretical treatment.
The regime where the coupling strength becomes comparable to the bare resonance frequencies in the system is  referred to as the ultra-strong coupling (USC) regime
~\cite{Forn2019}.
Previous works have established matrix-product state (MPS) simulations as a means to explore quantum optics phenomena of small atoms in the USC regime~\cite{Sanchez_Burillo_2014}, and they have proven useful for the study of non-Markovian light-matter interactions~\cite{Pichler_2016,Arranz_Regidor_2021}.
While the USC regime is becoming more and more experimentally accessible, its theoretical study still requires improved analytical and numerical methods, making it a timely research topic.
Moreover, at even stronger couplings and within the deep and extremely strong coupling regimes, other non-perturbative methods become available again~\cite{ashidaCavityQuantumElectrodynamics2021,ashidaNonperturbativeWaveguideQuantum2021}.

Here we investigate the low-energy physics and the dynamics of giant atoms beyond the RWA, in the USC regime and with multimode interactions, using a numerically exact, non-perturbative approach. 
We model the giant atoms as two-level systems.
The coupling points we model by a profile function with a finite width thus suppressing the coupling to high frequency modes and motivating a natural UV cutoff. Apart from this UV cutoff our approach requires no further approximations of the model Hamiltonian.
While our approach is general, we mainly focus on superconducting qubits coupled to acoustic field modes and the resulting non-Markovian effects which are due to a finite speed of sound.
In particular, we investigate the dynamics of a single giant atom coupled to an acoustic waveguide with intrinsic time delay, thus extending the analysis of previously predicted oscillating bound states~\cite{Guo2020} beyond the single-excitation subspace.
Our theoretical treatment of the system-reservoir interaction relies on a so-called chain (also star-to-chain or Lanczos) transformation.
This unitary transformation casts the field into the shape of a linear harmonic chain, which is particularly suited for numerical simulation. 
Rooting back to the numerical renormalization group~\cite{bullaNumericalRenormalizationGroup2008}, these methods are widely used in the study of open quantum systems (\eg see \cite{Chin2010,woodsSimulatingBosonicBaths2015,trivediConvergenceGuaranteesDiscrete2021}), but have also proven useful in quantum optics as seen, for example, in~\cite{busserLanczosTransformationQuantum2013,feiguinQubitphotonCornerStates2020,allerdtNumericallyExactApproach2019,Munoz2018}. 
We follow a similar numerical approach as \cite{Munoz2018}, which allows us to go beyond the single-excitation subspace and numerically study system and bath observables using MPS~\cite{Schollwoeck2011,Cirac2021}.

\begin{figure*}%
    \centering
    \includegraphics[width=.99\textwidth]{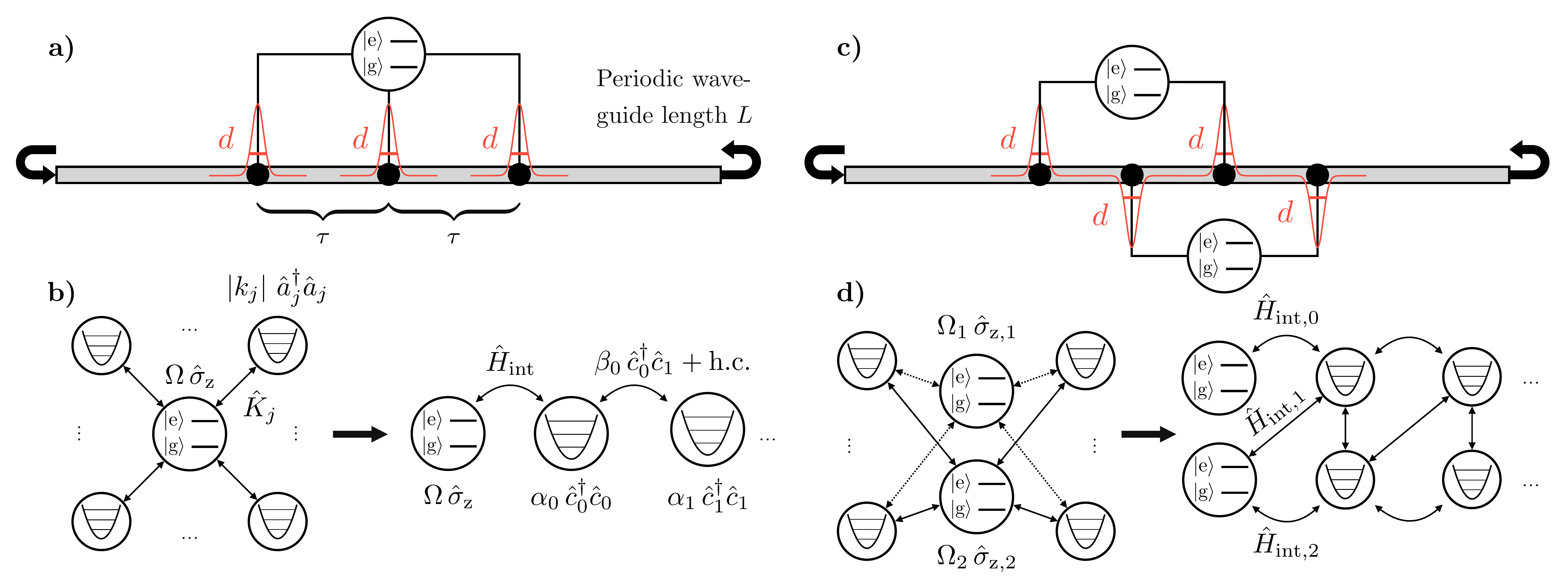}
    \caption{Setup and schematic illustration of chain transformation.
    (a) Giant atom with $M = 3$ equidistantly spaced coupling points at distance $\tau$, coupled to a periodic waveguide of length $L$.
    Emitter-waveguide couplings are locally described by Gaussian smearing functions $f^\mathrm{s}$ in Eq.~\eqref{eq:GaussianProfile}.
    (b) Chain transformation maps system-reservoir model with a one-to-all coupling to a linear chain as described by chain parameters $\alpha_{1, ..., N}$ and $\beta_{1, ..., N-1}$ in Eq.~\eqref{eq:chain-mapped-field}, with  $\hat K_j= \hat \sigma_{\mathrm{x}}\otimes ( f_j\hat a_j+f_j^*\hat a_j^\dagger )$.
    (c) Illustration of a two-atom setup with braided coupling points, which can be chain transformed into a chain with next-to-nearest-neighbor couplings, as indicated in (d).
    }
    \label{fig:setup}
\end{figure*}

This work is structured as follows.
In Sec.~\ref{sec:setup}, we introduce the setup and theoretical model of our study.
We show how the underlying Hamiltonian can be cast into a form amenable to an efficient numerical analysis even at strong coupling and beyond the RWA, using chain-mapping techniques.
In contrast to earlier works, our description does not rely on the assumption of a point-like emitter-bath coupling, but we promote coupling points to smeared coupling functions with a finite spatial support.
We provide estimates for the required values of all characteristic system parameters of an experimental implementation using superconducting circuits at the end of Sec.~\ref{sec:setup}.
In Sec.~\ref{sec:stationary}, we present an analysis of the low-energy physics of the system.
In particular, we discuss elementary excitations of the ground state as a function of increased emitter-reservoir coupling strength, in analogy with the well-understood quantum Rabi model.
In Sec.~\ref{sec:dynamic} we present a study of the temporal dynamics of a single giant atom coupled to an acoustic waveguide, with an intrinsic time delay, at three coupling points.
We showcase and discuss the implications of the breakdown of the RWA at strong coupling. A stability analysis of the findings with respect to experimentally relevant parameters is provided.
Finally, we conclude our work in Sec.~\ref{sec:outlook}, discuss possible future research directions and highlight the wide-ranged applicability of our approach, \eg to systems with multiple giant atoms and multilevel emitters. The latter is particularly important to realistic implementations in which, depending on the chosen gauge, the two-level approximation is no longer applicable for sufficiently strong couplings~\cite{debernardisBreakdownGaugeInvariance2018,stokesGaugeAmbiguitiesImply2019,rothOptimalGaugeMultimode2019}.
Note that we use natural units ($\hbar, c = 1$) throughout this work.

\section{Setup and theoretical framework \label{sec:setup}}
In this section, we present our theoretical framework and introduce the chain transformation that we employ for the study of stationary (see Sec.~\ref{sec:stationary}) and dynamical (see Sec.~\ref{sec:dynamic}) properties of two-level emitters coupled to a waveguide.

\textit{Setup}.\textemdash
A schematic illustration of the setup and the chain transformation is provided in Fig.~\ref{fig:setup}.
We treat a single quantum emitter as a two-level system coupled to the waveguide modes at $M$ coupling points, cf.~Fig.~\ref{fig:setup}(a).
For simplicity, we focus on equidistantly spaced coupling points, with a non-zero, significant propagation time $\tau$ between neighboring coupling points.
Such a system may be realized with a superconducting qubit piezoelectrically  coupled to an acoustic waveguide at several locations~\cite{Andersson2019}.
The interaction between emitter and waveguide modes is usually described by a one-to-all coupling,  \ie the emitter couples to all non-interacting field modes.
Once brought into the form of a linear chain, cf.~Fig.~\ref{fig:setup}(b), well-developed techniques based on MPS can be utilized for efficient numerical studies of various system and bath observables.
Note that for setups with multiple emitters, where $n$ emitters couple to one waveguide as schematically depicted in Fig.~\ref{fig:setup}(c), the chain transformation, as reviewed in App.~\ref{app:lanczos_math}, casts the field into a linear chain, with each mode coupling to its $n$ nearest  neighbors as indicated in Fig.~\ref{fig:setup}(d).

As mentioned, here we model the atom as a two-level system, \ie we use the two-level approximation (TLA).
For couplings above the weak coupling regime, the validity is known to be highly gauge dependent~\cite{debernardisBreakdownGaugeInvariance2018,stokesGaugeAmbiguitiesImply2019,rothOptimalGaugeMultimode2019} and only specific gauges still allow for the TLA to be applied beyond weak coupling. Here we chose the TLA interaction Hamiltonian akin to the dipole gauge which, for the quantum Rabi model was found to perform reasonably well in the USC regime~\cite{debernardisBreakdownGaugeInvariance2018}. 

\textit{Model}.\textemdash
The total Hamiltonian can be decomposed as the sum of the atomic, the field, and the interaction Hamiltonian,
\begin{equation}\label{eq:hamilt_tot}
    \Htot=\Ha+\Hf+\Hi.
\end{equation}
Assuming a two-level emitter with frequency $\Omega$, and a massless field in a periodic cavity of length $L$ described by modes with wavenumbers $k_j = 2\pi j/L$, the non-interacting terms in \eqref{eq:hamilt_tot} can be written as
\begin{align}
\Ha&=\frac\Omega2\left(\ketbra{\eup}\eup-\ketbra{\gup}\gup\right)=\frac\Omega2 \hat\sigma_\mathrm{z}, \\
\Hf&=\sum_{j} \left|k_j\right| \hat a_j^\dagger \hat a_j,
\end{align}
with the ground and excited states of the emitter, $\ket{\gup}$ and $\ket{\eup}$, and the annihilation (creation) operator $a_j^{(\dagger)}$ of field mode $j$.
The interaction Hamiltonian reads 
\begin{align}
    \Hi = \lambda \left(\ketbra{\eup}\gup+\ketbra{\gup}\eup\right)\otimes\integral{x}{}{} f(x)\hat\pi(x),
    \label{eq:h_int}
\end{align}
where $\lambda$ is a dimensionless coupling constant and $\hat \pi$ denotes the field momentum.
The smearing function $f(x)$ models the emitter-waveguide coupling and has the dimensions of a density.
For a giant-atom setup as shown in Fig.~\ref{fig:setup}(a), where $M>1$, we consider a sum of single-point couplings of the form
\begin{align}
    f(x)=\sum_{l=1}^M \pf(x-x_l),
\end{align}
with coupling points centered around the positions $x_1,...,x_M$.
The shape of $\pf(x)$ may not directly correspond to the physical shape of a given coupling point, but should be chosen to correctly capture the frequency dependence of the coupling strength (see Eq.~\eqref{eq:fj_defn}).
In this work, each coupling point is described by a Gaussian profile function
\begin{align}\label{eq:GaussianProfile}
    \pf(x)=\frac{\ee{-x^2/d^2}}{d\sqrt\pi},
\end{align}
with $2d$ being the effective diameter of each coupling point and $\integral{x}{}{} \pf(x)=1$.
Other choices for $\pf(x)$ can equally be considered, and some examples are discussed in App.~\ref{app:coupling_point_profiles}.
Note that the choice $\pf(x)=\delta(x)$ results in a UV divergent coupling which, however, does not occur in physical models~\cite{parra2018quantum}.

\begin{figure}%
    \centering
    \includegraphics[width=.48\textwidth]{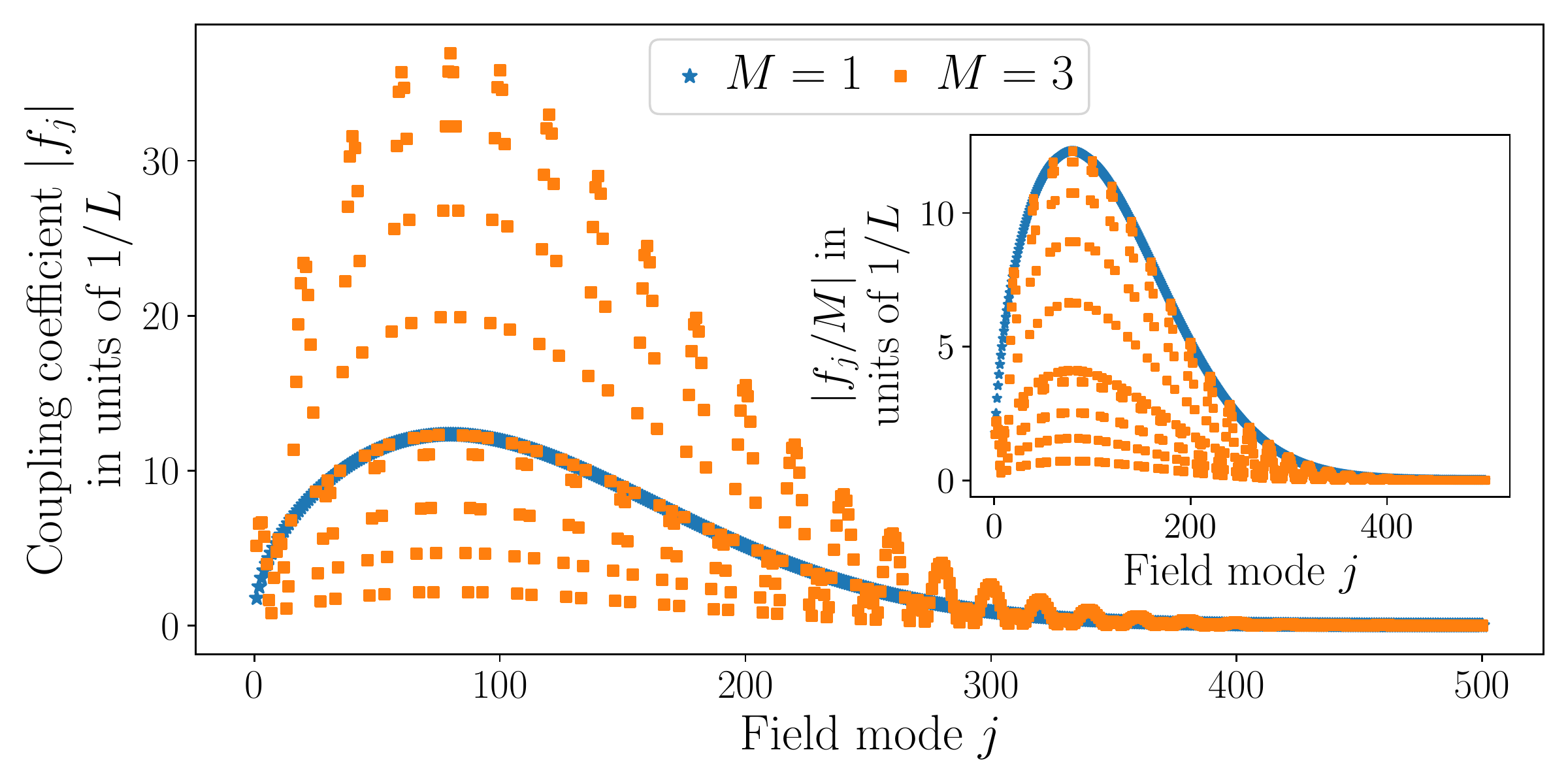}
    \caption{Coupling coefficients $|f_j|$ in~\eqref{eq:Gaussianfj} for an atom with a single coupling point $M=1$ (blue triangles), and a giant atom with $M=3$ as in~\eqref{eq:threepointfj}, spaced by $\tau=L/20$ (orange squares).
    }
    \label{fig:fij}
\end{figure}

\textit{Field modes}.\textemdash
The field momentum operator $\hat \pi(x)$, which is equal to the time derivative $\partial_t\hat\phi(x)$ of the field amplitude, expressed in terms of field eigenmodes, reads
\begin{align}
    \hat\pi(x)=-\ii \sum_j \sqrt{\frac{|k_j|}{2L}} \left(\ee{\ii k_j x}\hat a_j-\ee{-\ii k_j x}\hat a_j^\dagger\right).
    \label{eq:conj_momentum}
\end{align}
Hence, we can rewrite the interaction Hamiltonian as
\begin{align}
    \Hi &=\lambda\left(\ketbra{\eup}\gup+\ketbra{\gup}\eup\right)\otimes \sum_j f_j\hat a_j+f_j^*\hat a_j^\dagger , \\
    &f_j = -\ii \sqrt{\frac{|k_j|}{2L}} \integral{x}{}{} \ee{\ii k_j x} f(x).\label{eq:fj_defn}%
\end{align}
The coefficients $f_j$ for a giant atom with equidistant coupling points follow straightforwardly from the coefficients $f_j^\mathrm{s}=-\ii \sqrt{\frac{|k_j|}{2L}} \integral{x}{}{} \ee{\ii k_j x} \pf(x) $ for a single coupling point. 
For example, for a giant atom with three coupling points at $x_l=-\tau,0,\tau$ ($l = 1, 2, 3$), we find
\begin{align}\label{eq:threepointfj}
    f_j %
    =  \left(1+2\cos\left(k_j\tau\right)\right) \pf_j.
\end{align}
For the Gaussian profile~\eqref{eq:GaussianProfile}, replacing the integral $\integral{x}0L$ by $\integral{x}{-\infty}\infty$ since $d \ll L$, we obtain
\begin{align}\label{eq:Gaussianfj}
    \pf_j= -\ii \frac{\sqrt{j\pi}}L \ee{- \frac{d^2 \pi^2 j^2}{L^2}}.
\end{align}
The behavior of $|f_j|$ is shown in Fig.~\ref{fig:fij} for  emitters coupling to the waveguide with this smearing function through $M=1$ and $M=3$ points, respectively.

The decay of $|f_j|$ for sufficiently large $j$ allows us to introduce a UV cutoff and only consider a finite number of $2N$ field modes, \ie we restrict the index to $-N\leq j\leq N$ and also discard the zero mode, to which the atom does not couple.
Note that this UV cutoff is the only simplification of the original physical model that the present approach requires.
In particular, it does not rely on the rotating wave approximation (RWA) or the Wigner-Weisskopf approximation.

\begin{figure}%
    \centering
    \includegraphics[width=.48\textwidth]{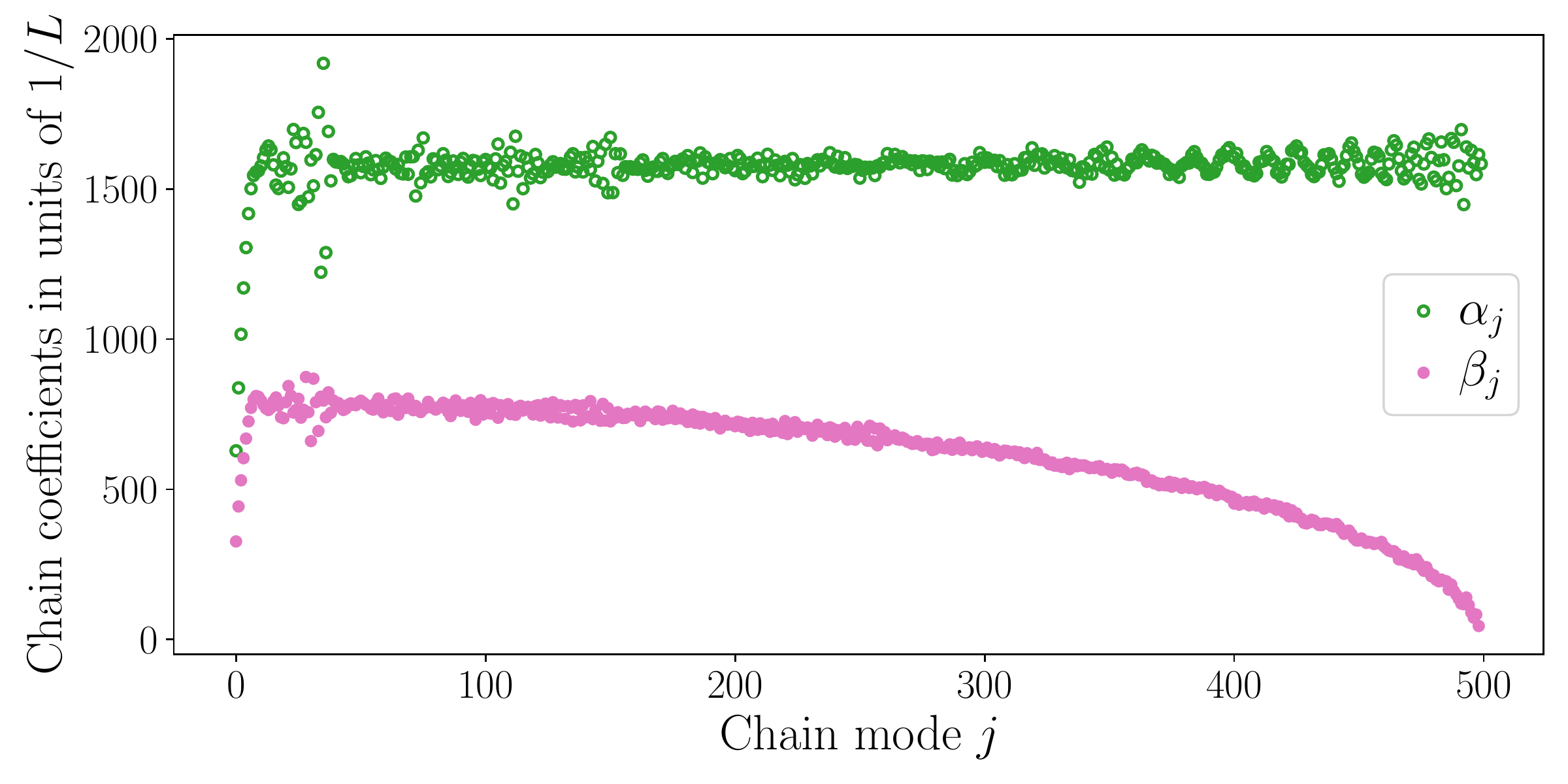}
    \caption{Coefficients appearing in the chain form of the field Hamiltonian~\eqref{eq:chain-mapped-field} for a giant atom with parameters as in Tab.~\ref{tab:default_values}.
    }
    \label{fig:chain_coefficients}
\end{figure}

\textit{Chain modes}.\textemdash
The dynamics of the Hamiltonian, after the UV cutoff, is straightforward to treat numerically if the coupling is weak, such that the RWA can be applied, and if one restricts attention to the single-excitation subspace of the approximate Hamiltonian.
However, in order to treat many excitations within the RWA, and to study USC beyond the domain of the RWA, here we employ a chain transformation of the field modes.
Such a chain transformation yields a new basis of field mode operators $\hat c_0,...,\hat c_{2N-1}$, which we refer to as chain modes. These are related to the eigenmodes of $\Hf$ by a non-mixing Bogoliubov transformation,
\begin{align}
    \hat c_i=\sum_{j=-N}^N \Lambda_{ij}\hat a_j.
\end{align}
The front chain mode is chosen as $\hat c_0=\frac1{\sqrt{\mu_0}}\sum_j f_j \hat a_j$ with
\begin{align}
    \hat c_0=\frac1{\sqrt{\mu_0}}\sum_j f_j \hat a_j,\quad \mu_0=\sum_j\left|f_j\right|^2,
\end{align}
such that the interaction Hamiltonian takes the form
\begin{align}\label{eq:chain-mapped-interaction}
    \Hi&=\lambda\sqrt{\mu_0}\left(\ketbra{\gup}\eup+\ketbra{\eup}\gup\right)\otimes \left(\hat c_0+\hat c_0^\dagger\right).
\end{align}
Using Lanczos algorithms (see App.~\ref{app:lanczos_math})  the chain modes are appropriately chosen such that they cast the field Hamiltonian into the form of a harmonic chain with nearest-neighbor hopping interactions only,
\begin{align}\label{eq:chain-mapped-field}
    \Hf=\sum_{i=0}^{2N-1} \alpha_i\hat c_i^\dagger\hat c_i +\beta_i \left(\hat c_i^\dagger \hat c_{i+1}+\hat c_{i+1}^\dagger\hat c_i\right),
\end{align}
with real coefficients $\alpha_i,\beta_i\in\RR$. Fig.~\ref{fig:chain_coefficients} shows a plot of these coefficients for the setup that we will use in our numerical examples (see Tab.~\ref{tab:default_values}).

Note that if the atom has an even profile function $f(x)=f(-x)$ such as~\eqref{eq:threepointfj}, it does not couple to the odd sector of the field modes.
Then, by introducing the basis change $\hat a^{(\pm)}_j=\left(\hat a_j\pm\hat a_{-j}\right)/\sqrt2$, we can restrict attention to the $N$ field modes of the even sector and,
accordingly, only construct $N$ chain modes as linear combinations of even field modes.

\textit{Numerical simulations}.\textemdash
The low-energy sector of \eqref{eq:hamilt_tot} can be efficiently described using MPS~\cite{Schollwoeck2011,PAECKEL2019167998}, once the interaction and field Hamiltonians have been cast into their respective forms \eqref{eq:chain-mapped-interaction} and \eqref{eq:chain-mapped-field}.
In the following sections, we use both density-matrix renormalization group (DMRG) and time-evolution algorithms to study the stationary and dynamical properties of giant atoms coupled to a waveguide, as a function of the coupling strength $\lambda$ and the emitter frequency $\Omega$.

\begin{table}[t]
    \centering
    \renewcommand{\arraystretch}{1.5}
    \begin{tabular}{ p{1.2cm} p{4.1cm} p{2.8cm}} 
        \toprule
         {\bf symbol} & {\bf giant atom property} & {\bf default value}   \\
        \colrule
         & waveguide free spectral range & $2\pi/L$ \\
        $M$ & number of coupling points &  $M=3$    \\
        $\tau$ &  coupling point distance & $\tau=L/20$ \\
        $d$ & width of Gaussian profile~\eqref{eq:GaussianProfile} & $d=L/500$ \\
         $\sqrt{\mu_0}$ & interaction energy scale~\eqref{eq:chain-mapped-interaction} & $\sqrt{\mu_0}\approx 345.1/L$\\
        $\lambda$ & coupling strength & $\lambda=0.4$ \\
        $\Omega$ & atom frequency & $\Omega=|k_{80}|=160\pi/L$\\
        \botrule
    \end{tabular}
    \caption{Giant atom geometry and parameters used as default in figures and numerical results, unless stated otherwise. The (periodic) waveguide's length $L$  sets the overall length scale.}
    \label{tab:default_values}
\end{table}

Unless stated otherwise, the default configuration  that we consider is that of a giant atom coupling to the chain modes  at $M=3$ coupling points, located at $x_l=-L/20,\,0,\,L/20$, each modelled by the Gaussian smearing~\eqref{eq:GaussianProfile}, with all parameters as specified in Tab.~\ref{tab:default_values}.
The default value for the atom frequency $\Omega$ is chosen to be resonant with the 80\textsuperscript{th} field mode which features the largest coupling coefficient $|f_j|$, as can be seen in Fig.~\ref{fig:fij}.

At strong coupling $\lambda$, care must be taken to ensure that the truncation errors associated with increasing chain-mode occupation numbers $\langle \hat c^\dagger_i \hat c_i \rangle$ are still negligible in the numerical calculations.
We find that this is possible even deep in the USC regime, as discussed in Sec.~\ref{sec:stationary}, using $25$ bosons per site.
In our numerical calculations using the iTensor software package~\cite{itensor}, we also ensure convergence with respect to MPS bond dimension ($\approx 200$) and chain length $N \leq 1000$ at a singular value decomposition (SVD) cutoff of $10^{-12}$.

\textit{Experimental considerations.}\textemdash
Experimentally, the present system and its considered initial state can be realized and prepared, \eg using superconducting qubits coupled to an acoustic cavity~\cite{Manenti2017,Moores2018,Chu2018,Andersson2019}.
Based on prototypical parameters used in our calculations, cf.~Table~\ref{tab:default_values},  one may choose a qubit frequency of $\Omega/(2\pi)=2.4$~GHz.
At a typical sound velocity of $c = 3~$km/s this yields a distance of~$\approx 5~\mu$m between the coupling points of the atom.
In an implementation, instead of a periodic waveguide with length~$L$, one can consider an open-ended waveguide of length~$L/2\approx 50~\mu$m.
These ballpark values are realistic and consistent with recent experimental implementations.
As described in Secs.~\ref{sec:stationary} and~\ref{sec:dynamic}, we identify the onset of USC at around $\lambda \approx 0.15$, which amounts to an acoustic qubit-waveguide coupling of~$\approx 8.5~$MHz per coupling point.
Considering that the qubit couples through three coupling points, the total coupling between qubit and waveguide is comparable to the estimated  free spectral range of~$\approx 30~$MHz, placing the setup in the strong-multimode regime.
This value is comparable with previously reported acoustic coupling strengths, and there are various prospects for state-of-the-art experimental settings to be operated even more deeply in the strong coupling regime, \eg by appropriate choice of material~\cite{Moores2018}.

\section{Low-energy spectrum and eigenstates \label{sec:stationary}}

With the approach presented above, it is possible to investigate giant atoms beyond the realm of the RWA and the single-excitation subspace.
Since the field is not traced out for an effective-system description, the approach also yields full access to field observables such as photon numbers or field energy density, and allows for, \eg the investigation of virtual photon clouds.
As a first step,  here we calculate and characterize the ground state and first excited state of our giant atom setup, as a function of the coupling strength.
Hereby, we explore the entire USC regime and access the onset of the deep-strong coupling (DSC) limit.

For single-mode models, such as the QRM, the USC and DSC regimes are  well understood and characterized~\cite{RN114}, and both have been achieved on several experimental platforms.
In the following, we will see that the lowest energy eigenstates of our multimode model generally follow the intuition based on the single-mode QRM.
Yet the onset of signatures related to the USC and DSC regimes is shifted to smaller coupling strengths, which underlines that the effective emitter-field coupling is enhanced as the emitter simultaneously couples to the field via a multitude of modes.

\begin{figure}%
    \centering
    \includegraphics[width=.48\textwidth]{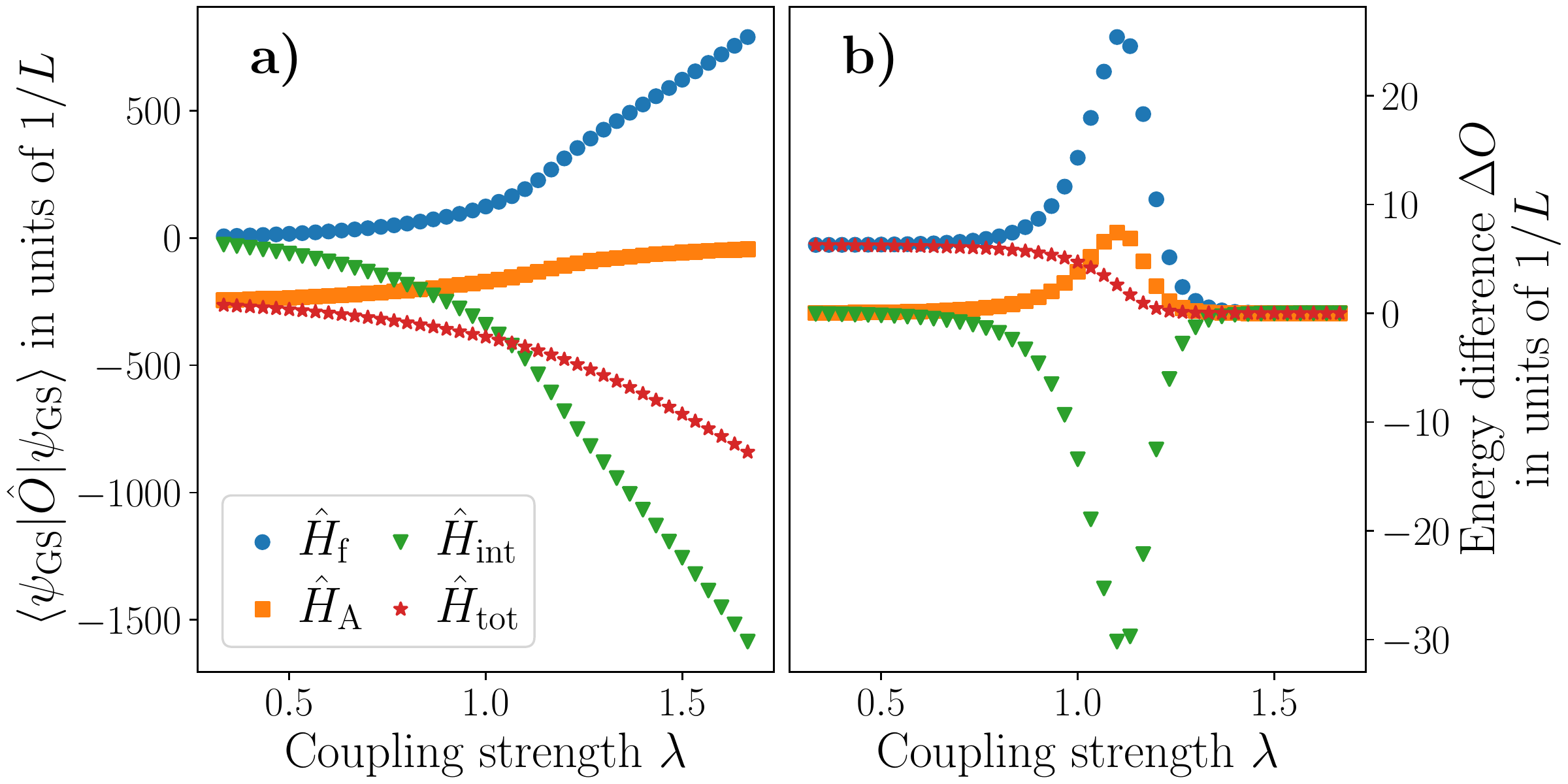}
    \caption{Lowest eigenenergies of $\hat H$ as function of coupling $\lambda$.
    (a) Ground-state energy $\langle \hat H \rangle$ (red, triangle) and contributions from atomic Hamiltonian $\langle\Ha\rangle$ (orange, triangle), field Hamiltonian $\langle\Hf\rangle$ (blue, circle) and interaction Hamiltonian $\langle\Hi\rangle$ (green, triangle).
    (b) Gap $\Delta H$ between ground and first-excited state energies (red, triangle), and decomposition into contributions as in (a).
    The labels are the same in (a) and (b).}
    \label{fig:stationary-energies}
\end{figure}

\textit{Eigenenergies}.\textemdash
Once the Hamiltonian $\Htot$ in~\eqref{eq:hamilt_tot} is transformed into a chain, we compute its ground state $|\psi_\mathrm{GS}\rangle$ and its  first-excited state $|\psi_\mathrm{ES}\rangle$ using DMRG.
In Fig.~\ref{fig:stationary-energies} we show the obtained ground-state energy $\langle\psi_\mathrm{GS}| \hat O |\psi_\mathrm{GS}\rangle$ 
and the difference $\Delta O = \langle\psi_\mathrm{ES}| \hat O |\psi_\mathrm{ES}\rangle - \langle\psi_\mathrm{GS}| \hat O |\psi_\mathrm{GS}\rangle$, for the total Hamiltonian ($\hat O=\Htot$) as well as separately for the atom ($\hat O=\Ha$), the interaction ($\hat O=\Hi$) and the field ($\hat O=\Hf$), as a function of the coupling strength $\lambda$.

The absolute values for the ground state in Fig.~\ref{fig:stationary-energies}(a) behave monotonically and, thus, make it difficult to distinguish different regimes.
However, the energy differences plotted in Fig.~\ref{fig:stationary-energies}(b) provide a richer picture: The energy gap of the Hamiltonian $\Delta H_\mathrm{tot}$ starts at $\Delta H_\mathrm{tot}(\lambda=0) =|k_1|=2\pi/L$ for $\lambda=0$, then decreases over an intermediate range of $0.5\lesssim \lambda\lesssim 1.5$, and, finally,  closes at $\lambda\gtrsim 1.5$.
The energy differences for the separate terms of the Hamiltonian behave accordingly at low and large $\lambda$, but they exhibit prominent peaks in the intermediate region, where the energy gap $\Delta H_\mathrm{tot}$ is closing most rapidly.
The behavior of the spectrum in the intermediate range of $\lambda$ resembles the spectrum of the single-mode QRM~\cite{RN114} in the USC, whereas for $\lambda\gtrsim1.5$, the spectrum resembles the DSC of the QRM.
In the QRM, the USC sets on when the ratio of coupling strength to emitter gap is of the order of $\sim 0.1$, and DSC sets on at a ratio of $\sim 1$.
In our approach, analogously, the range of USC can be estimated by considering the ratio of the energy scale of the interaction Hamiltonian~$\lambda\sqrt{\mu_0}$ to the atom's gap~$\Omega$, where we have $\sqrt{\mu_0}/\Omega\approx 0.69$.
From this comparison, one expects the USC to lie within $0.15\lesssim \lambda\lesssim 1.5$, which agrees well with our numerical findings.
In contrast, the coupling strength between emitter and the resonant field mode is only $|f_{80}|/\Omega\approx 0.076$, which wrongfully would suggest the USC regime to lie at much higher $\lambda$.
This underlines that the emitter couples efficiently to many field modes, cf.~Fig.~\ref{fig:fij}, and that a single-mode description would fail.

\textit{Structure of eigenstates}.\textemdash
We can further investigate the structure of the obtained lowest eigenstates in the different coupling regimes, and compare them to our expectations based on the QRM, by  characterizing them in terms of observables such as photon numbers and the atomic population.

\begin{figure}%
    \centering
    \includegraphics[width=.48\textwidth]{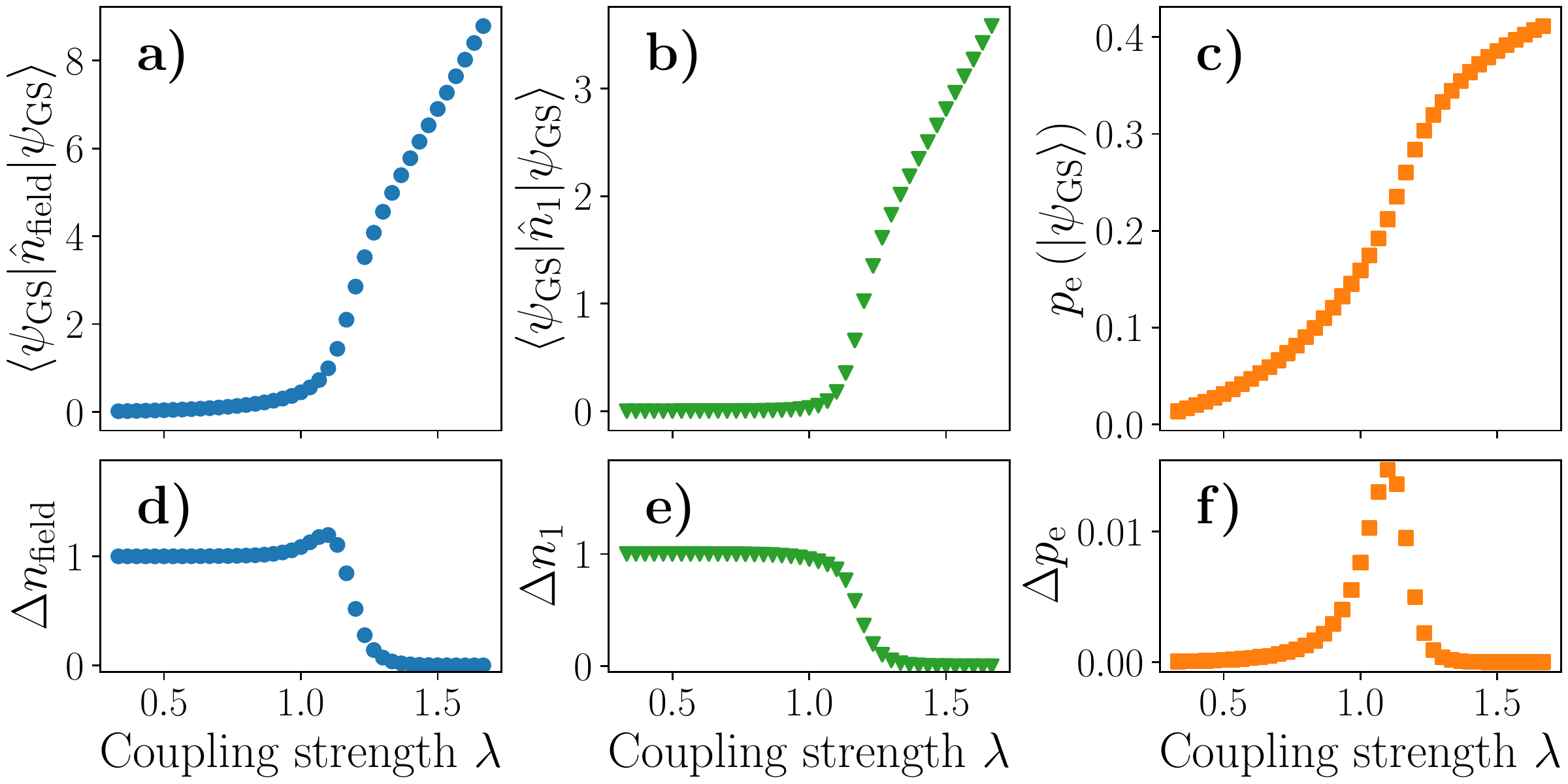}
    \caption{
    Atomic population and occupation of field modes in the ground and first excited state, as a function of coupling strength $\lambda$. The upper row shows (a) the expectation values of the total photon number $\hat n_{\mathrm{field}}$, as well as (b) the number operator of the first field mode $\hat n_1=\hat a^\dagger_1\hat a_1$, and of (c) the atomic occupation.
    Analogously,  (d)--(f)  show the increase of the expectation value in the first excited state, \ie  $\Delta n_1 = \langle \psi_\mathrm{ES}|\hat n_1|\psi_\mathrm{ES}\rangle-\langle\psi_\mathrm{GS}|\hat n_1|\psi_\mathrm{GS}\rangle$, \textit{etc.}
    }
    \label{fig:stationary-occupations}
\end{figure}

In the perturbative regime, as $\lambda\to0$, clearly the ground state of the system approaches the free ground state, \ie the product of the atom ground state and the vacuum $\ket{\psi_{\mathrm{GS}}}\to\ket{\gup,0}$, and the first excited state is obtained by placing a single photon into the first free field mode $\ket{\psi_{\mathrm{ES}}}\to\hat a^\dagger_1 \ket{\psi_{\mathrm{GS}}}$.
In fact, as shown in Fig.~\ref{fig:overlap} of App.~\ref{app:overlap}, we find a large overlap $|\bra{\psi_{\mathrm{ES}}} \hat a^\dagger_1 \ket{\psi_{\mathrm{GS}}}|^2$ for our numerically obtained eigenstates for sufficiently small coupling.

To characterize the eigenstates in USC and beyond, we consider the atomic population $p_\mathrm{e} = (1+\langle \hat \sigma_\mathrm{z} \rangle)/2$,
the total excitation number of the field $\langle\hat n_{\mathrm{field}}\rangle=\sum_j\langle\hat n_j\rangle$,
with $\langle \hat n_j \rangle = \langle \hat a_j^\dagger \hat a_j \rangle$ for the $j$th field mode,
and the occupation number of the lowest free field mode $\langle \hat n_1\rangle$.
Figs.~\ref{fig:stationary-occupations}(a)-(c) show these expectation values in the ground state, and their increase as a function of coupling strength $\lambda$.
The lower panel, Fig.~\ref{fig:stationary-occupations}(d)-(f), displays the difference $\Delta O = \langle \psi_\mathrm{ES} | \hat O | \psi_\mathrm{ES} \rangle - \langle \psi_\mathrm{GS} | \hat O | \psi_\mathrm{GS} \rangle$  for each observable $\hat O$.

For the lowest values of $\lambda$, we recognize the  results of  $\ket{\psi_{\mathrm{GS}}}$ and $\ket{\psi_{\mathrm{ES}}}$ lying close to $\ket{\gup,0}$ and $\hat a_1^\dagger\ket{\gup,0}$, respectively. For large couplings, where we saw above that the two pairs approximately form a degenerate pair, we see that the states also agree in the occupation observables.
This pair is characterized by a large number of field excitations that are spread out over many field modes, and by the atom approaching half occupation, $p_\mathrm{e} \rightarrow 1/2$.
In fact, this is what one would expect in the DSC where the interaction Hamiltonian $\Hi$ dominates over the other parts of the total Hamiltonian $\Htot$ and, thus, eigenstates of $\Htot$ lie close to eigenstates of $\Hi$.
Eigenstates of $\Hi$, however, would be given by the product of the eigenstates of $\hat\sigma_{\mathrm{x}}=\ketbra{\eup}\gup+\ketbra{\gup}\eup$, given by $\ket{\pm X}$, and  eigenstates of the field operator $\hat c_0+\hat c_0^\dagger$,
which can be approximated well by coherent states with a (positive or negative) eigenvalue with respect to $\hat c_0$, thus allowing for the construction of a degenerate pair.
This behavior is reminiscent of the eigenstates of the single-mode QRM. Also there, for small coupling, the ground state of the system contains no photonic excitations, and in the DSC regime, the number of photonic excitations grows linearly while the atomic population saturates at half occupation.

\section{Oscillating bound states \label{sec:dynamic}}
\begin{figure}
    \centering
    \includegraphics[width=.48\textwidth]{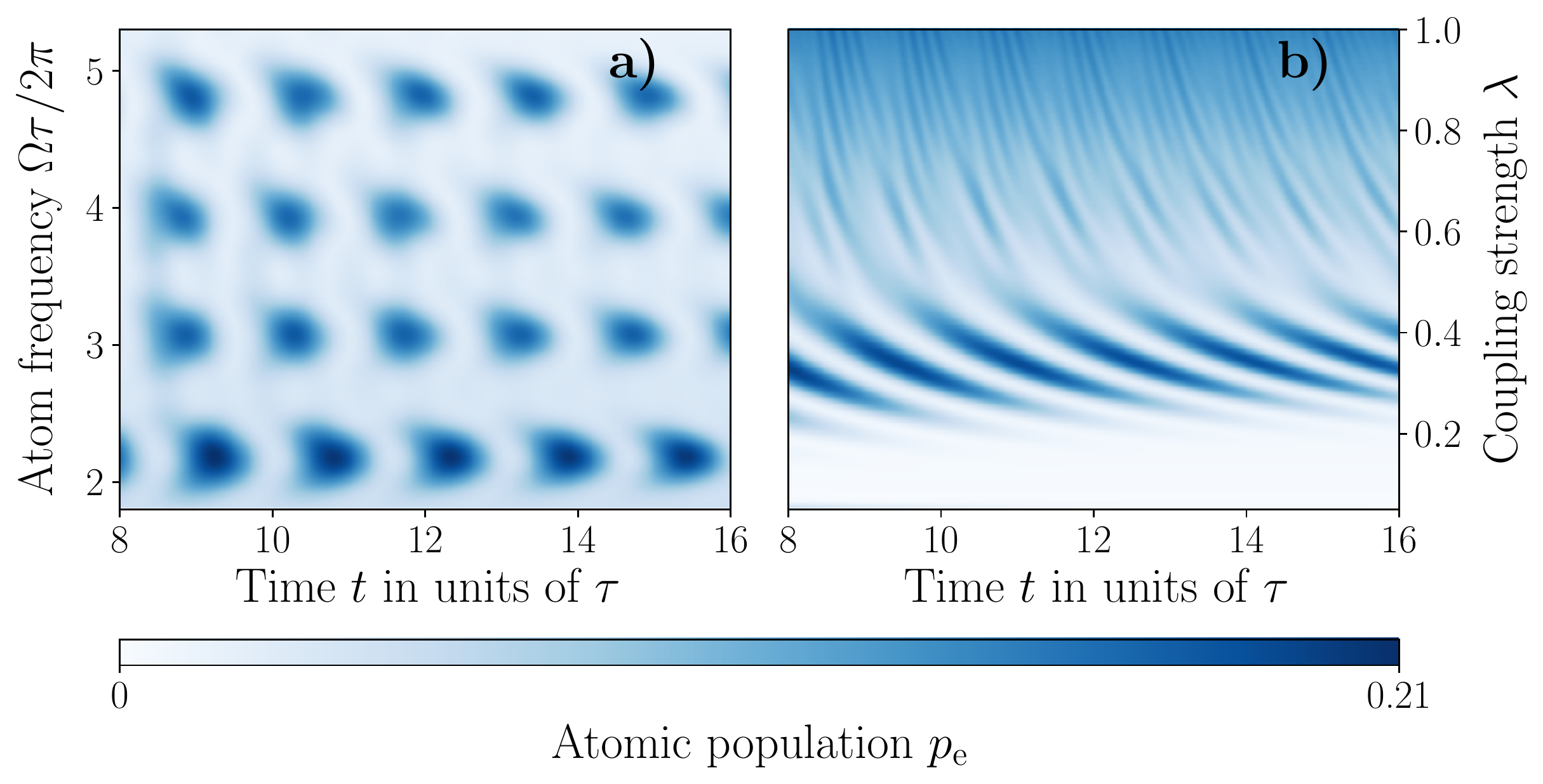}
    \caption{Dynamic of the atomic population for variation (a) of the atom frequency $\Omega$ and (b) of the coupling strength $\lambda$, whereas the other parameter is kept at its default value of (a) $\lambda=0.4$ and (b) $\Omega\tau/(2\pi)=4$. The atom starts excited with $p_e=1$ at $t=0$. 
    For certain parameter regions  oscillating bound states form.
    Note that the coupling point separation $\tau=L/20$ is used as the length scale.
    (Step sizes used for plotting: time $\Delta t = 5\times 10^{-4}$, frequency $\Delta \Omega$ = 0.04, coupling strength $\Delta \lambda$ = 0.0025.)
    }
    \label{fig:oscillating_omega_lambda}
\end{figure}

This section studies dynamical properties of giant atoms with a focus on  oscillating bound states.
These states were recently predicted to arise for giant atoms in the RWA regime under certain, fine-tuned conditions.
In our approach we can simulate the dynamics of giant atoms far into the USC regime, and up to times set by the waveguide crossing time, before finite-size effects occur.
Here we observe the formation of oscillating bound states and show that they are robust against variations  in the coupling parameters.

Oscillating bound states are a fascinating phenomenon of giant atoms:
When an initially excited giant atom decays into a waveguide then, under certain resonance conditions, 
a significant part of the energy may end up oscillating back and forth between the atom and  field.
The first derivation of this phenomenon in~\cite{Guo2020}, using RWA and $\delta$-coupling points for the atom,
identified specific combinations of parameter values for  the number of coupling points, the coupling strength and the atom's frequency, at which oscillating bound states appear.
In particular, in view of future experimental studies, this raises the questions of whether oscillating bound states can also be expected for finite-width coupling points, whether the appearance of oscillating bound states is robust against deviations in the coupling and frequency parameters, and whether oscillating bound states also appear in the strong coupling regimes.
In the following, we are able to answer these questions in the affirmative.

\begin{figure}
    \centering
    \includegraphics[width=.48\textwidth]{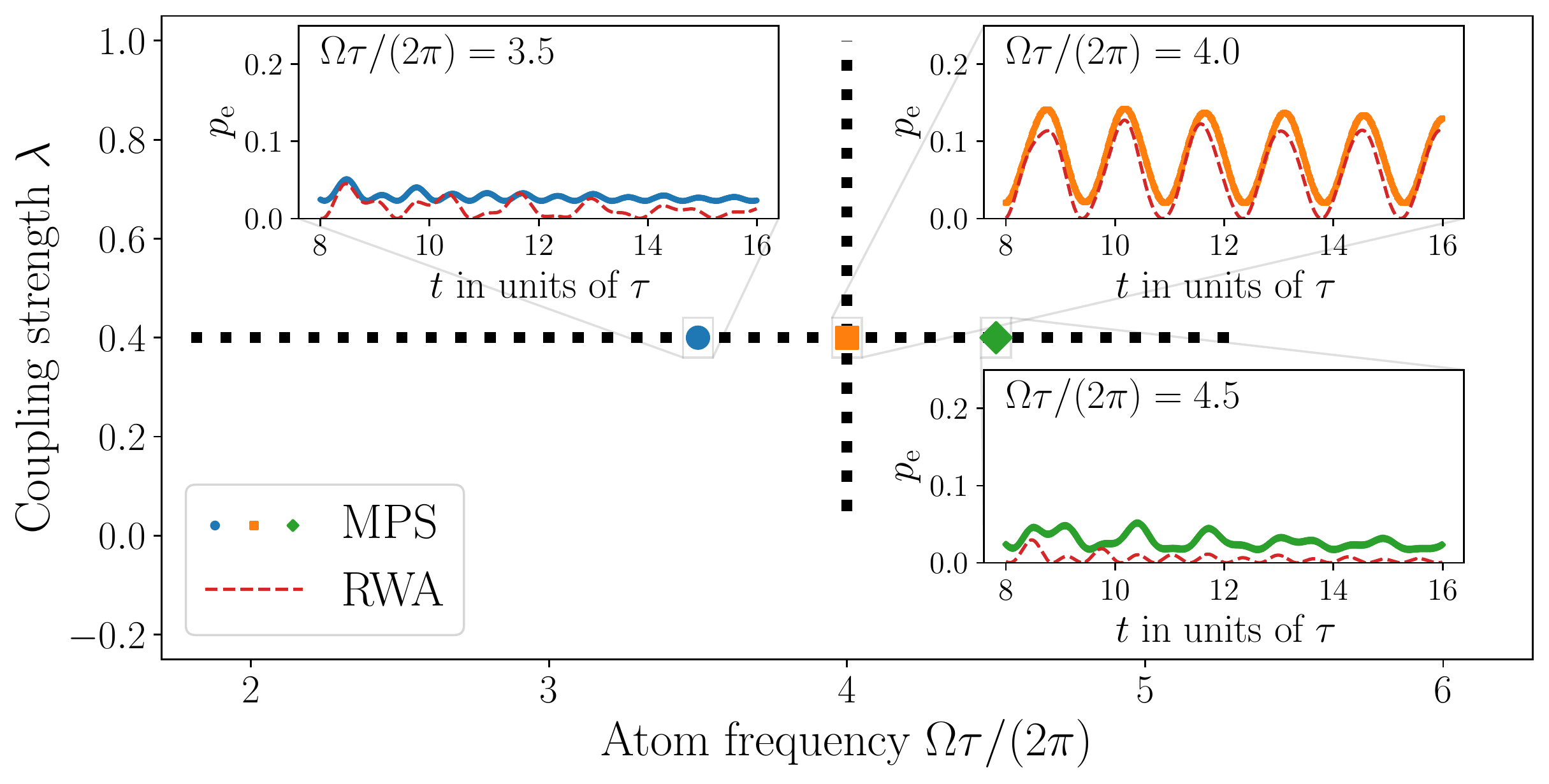}
    \caption{Visualization of the $\Omega-\lambda$-parameter space explored in Fig.~\ref{fig:oscillating_omega_lambda}, together with a comparison of the exact MPS results to RWA calculations. The dotted black lines represent the parameters plotted in Figs.~\ref{fig:oscillating_omega_lambda}(a) and~(b), with their intersection point corresponding to the default parameters of Tab.~\ref{tab:default_values}.
    The insets compare results from Fig.~\ref{fig:oscillating_omega_lambda} (solid lines) with results obtained using RWA (dashed, red lines). %
    }
    \label{fig:fadenkreuz}
\end{figure}

Fig.~\ref{fig:oscillating_omega_lambda} demonstrates the appearance of an oscillating bound state for the giant atom setup as introduced in Tab.~\ref{tab:default_values}, whose parameters were chosen  to correspond closely to an oscillating bound state configuration of~\cite{Guo2020}. 
The giant atom is initially in the excited state $\ket{\eup}$ when it is coupled to the waveguide in the vacuum at time $t=0$, \ie it starts out with a population of $p_e=1$.
After the initial decay process, which takes of the order of approximately $5\tau$ to $10\tau$, the system can realize an oscillating bound state, for certain parameters.
These states are characterized by a steady oscillation in the atomic population.
Because  our setup uses a periodic waveguide, we can only meaningfully describe the atom's dynamics up to  $t\lesssim 18\tau$.
After this time, radiation emitted at $t=0$ has traversed the waveguide and reaches back to the atom from the other direction
\bibnote[note_num_costs]{For the Trotterized time evolution of MPS in Figs.~\ref{fig:oscillating_omega_lambda} and \ref{fig:fadenkreuz}, we truncate the coupled oscillators of the chain (see Fig.~\ref{fig:setup}) using two bosons per site.
    At $\lambda \leq 0.6$,  we use a truncation error cutoff of $10^{-9}$, and for $\lambda > 0.6$, a cutoff of $10^{-8}$.
    Using these parameters, it took approximately $1800$ CPU hours to generate Fig.~\ref{fig:oscillating_omega_lambda}(a) and $800$ CPU hours to generate Fig. \ref{fig:oscillating_omega_lambda}(b).
}.

The plots of Fig.~\ref{fig:oscillating_omega_lambda} suggest that the appearance of oscillating bound states, to a certain extent, is robust against variations  both in the atom frequency and the coupling strength. 
In view of the fact that our approach accounts  for a non-zero, realistic width of the coupling points, this observation appears encouraging with respect to experimental implementations.
Fig.~\ref{fig:oscillating_omega_lambda}(a), where the atom frequency $\Omega$ is varied while the coupling strength is fixed at $\lambda=0.4$, shows regions with oscillating bound states appearing roughly periodically.
Fig.~\ref{fig:oscillating_omega_lambda}(b), where the atom frequency is fixed at $\Omega=160\pi/L$ while the coupling strength is varied, shows oscillating bound states only in the range of $0.3\lesssim\lambda\lesssim0.4$.
Within the RWA~\cite{Guo2020}, one expects oscillatory bound states to appear periodically both in $\Omega$ and in $\lambda$. However, based on the analysis in the previous section, we would count all data in Fig.~\ref{fig:oscillating_omega_lambda}(a), and all data in Fig.~\ref{fig:oscillating_omega_lambda}(b) with non-trivial dynamics, towards the USC regime, and thus beyond the regime where the RWA is valid.

Fig.~\ref{fig:fadenkreuz} demonstrates that the parameter regime we consider requires numerically exact calculations, by comparing our MPS results to calculations obtained  within the RWA.
It also illustrates that the error introduced by the RWA can change unexpectedly, probably due to the multimode couplings of our approach.
At the default values of $\Omega\tau/(2\pi)=4.0$ and $\lambda=0.4$, the agreement between the RWA and MPS results may still appear acceptable. 
Estimating the dimensionless emitter-waveguide coupling by $\lambda\sqrt{\mu_0}/\Omega$, one may thus assume that the agreement between the MPS and RWA calculations will improve when $\lambda$ is decreased or $\Omega$ is increased.
As far as the coupling strength is concerned, this is what we observe.
However, in the atom frequency, the agreement of the RWA and MPS results is highly non-monotonous.
Whereas there is an overall trend for the agreement to improve as $\Omega$ is increased, significant oscillations in the quality of the approximation can be observed.
In Fig.~\ref{fig:fadenkreuz}, this is illustrated by the inset for the population dynamics at $\Omega \tau/(2\pi) = 4.5$.
Here, in contrast to the  reasonably good agreement  between both curves in the presence of the oscillating bound state at $\Omega\tau/(2\pi)=4$, the RWA results in a significantly different prediction for the atom population.

The results of this section show that our approach allows one to explore the dynamics of the system far into the USC regime.
In fact, our numerical results indicate that for the setup considered here, intermediate-time evolutions are feasible up to a coupling strength of $\lambda\approx1.5$, at which point  (i) the MPS simulations become too costly and (ii) the system enters the DSC regime, as outlined in Sec.~\ref{sec:stationary}.
Thus, since other, perturbative approaches are more suitable at DSC, we expect our method to be most useful in the intermediate USC regime.
Further improvements of our numerical approach can be made by careful choice of parameters, \eg cavity length $L$.
Increasing  $L$ in order to extend the maximal simulation times, or to decrease the free spectral range of the cavity, would result in a larger number of modes in the field below the UV cutoff that need to be taken into account.  However, since the resulting chain length scales linearly in the cavity length, the increase in computational costs (cf.~\bibnotemark[note_num_costs]) may well be feasible.
 This could prove useful for further investigations of systems, \eg motivated by concrete experimental setups.

\section{Conclusions \& Outlook \label{sec:outlook}}

In summary, we have investigated the low-energy sector and time dynamics of a giant atom coupled to a waveguide, beyond the rotating-wave approximation.
We have outlined in detail how a prototypical model describing a giant atom coupled to all non-interacting field modes below a physically well-motivated UV cutoff can be conveniently cast into a form which is amenable to efficient numerical treatment using matrix-product states.
This approach has enabled us to compute the low-energy spectrum of the system at highly elevated coupling strengths, \ie going beyond the single-excitation subspace and identifying the onset of different strong-coupling regimes.
Based on previous findings in the context of the thoroughly explored standard quantum Rabi model, we have identified these regimes as ultra-strong and deep-strong coupling limits.
In contrast to earlier work, we have described the coupling between emitter and waveguide not as a point-like coupling, but using a profile function with a finite spread, suppressing the coupling to high-frequency field modes and allowing for a UV cutoff.
Since the presented approach is numerically exact and provides full access to a variety of system and bath observables, we were able to analyze how the contributions to the ground and first-excited state energies are distributed among the system, field and interaction Hamiltonians.
Using our numerical toolbox, we have calculated the low-lying eigenstates with up to $\approx 10$ excitations in the entire system, including bath and emitter.
Based on the relatively low computational costs of these simulations, our study paves the way for further numerical investigations of waveguide quantum electrodynamics with multiple giant atoms in all coupling regimes.
Furthermore, we have studied the time evolution of the composite system in an acoustodynamical setting, which may be realized by coupling a superconducting qubit at several locations to an acoustic waveguide.
It has previously been suggested that such setups, when operated in the non-Markovian limit, can host bound states characterized by a persistent exchange of energy between the artificial atom and its environment.
Here we have explicitly taken into account the significant time delay caused by a finite propagation speed of the acoustic modes, to investigate the pronounced non-Markovian features that arise as a consequence.
In contrast to earlier works, we have not been restricted to the single-excitation subspace, and demonstrated the emergence and robustness of oscillating bound states over a wide parameter range.
In particular, the breakdown of the rotating-wave approximation can be carefully monitored by applying our ansatz to the models with and without counter-rotating terms, respectively.

\begin{figure}
    \centering
    \includegraphics[width=.48\textwidth]{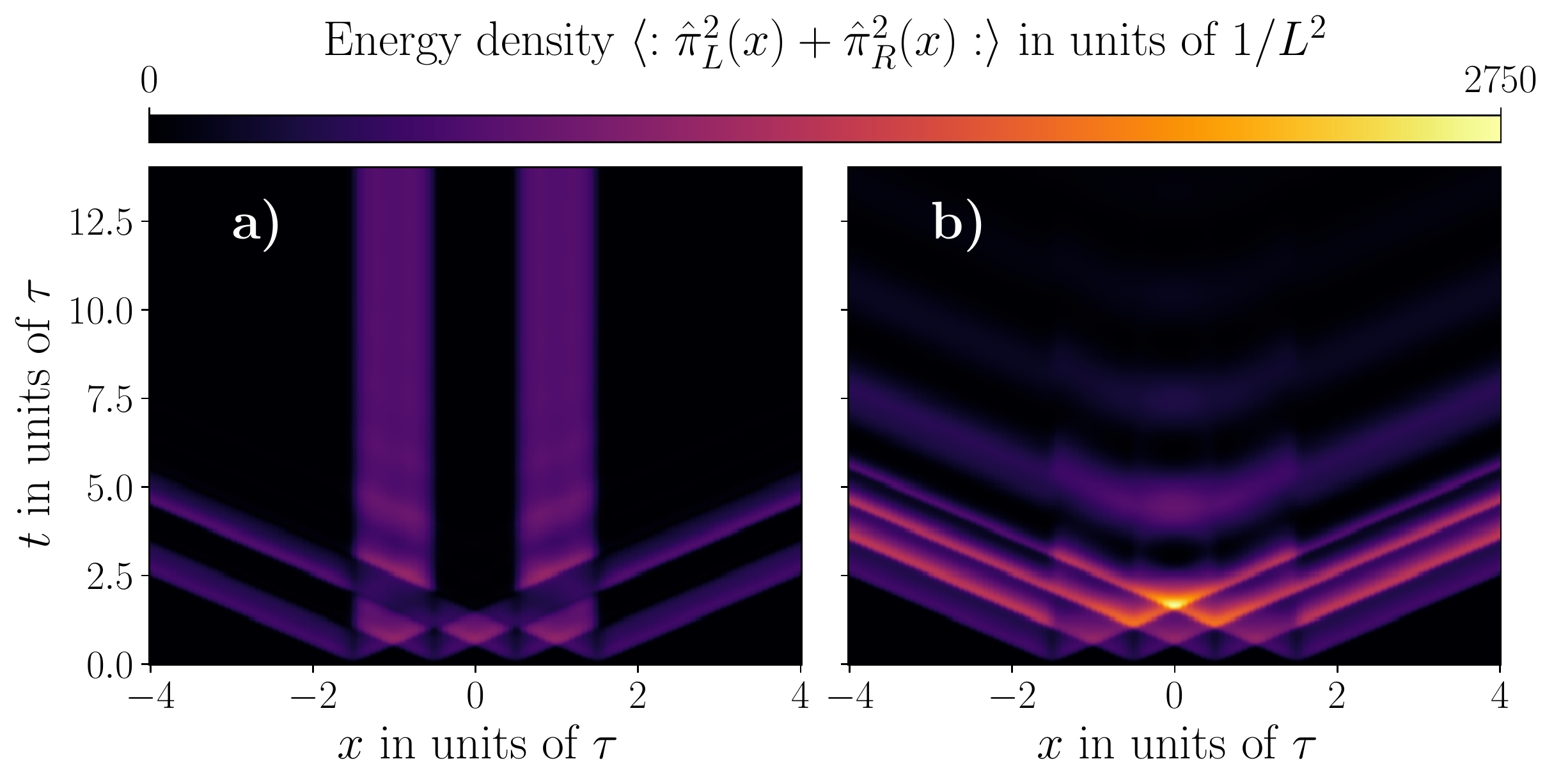}
    \caption{Energy density emitted by two giant atoms, as in Fig.~\ref{fig:setup}(c), initialized in entangled Bell states, in (a) for triplet and in (b) the singlet state, as the initial state.
    Each atom has two coupling points which are arranged in a braided configuration, \ie neighboring points are separated by~$\tau / 2$ and belong to alternating atoms.
     $\Omega_1\tau / (2\pi) =\Omega_2\tau / (2\pi) = 5.0, \lambda_1=\lambda_2 = 0.208, \tau = L/10,  d = L / 300$.
     (For details, see App.~\ref{app:energy_density}.)}
    \label{fig:energy_density_two_giant_atoms}
\end{figure}

Beyond the scope of this work, which focuses on single giant atoms, the chain transformation approach opens up the opportunity to also non-perturbatively study systems composed of two or several giant atoms coupled to a common environment, and within the ultra-strong coupling regime.
In fact, already within the rotating-wave approximation, it can offer advantages since simulations of chain transformed systems, based on matrix-product states, can treat many numbers of excitations in the system without any adjustments, whereas the Hilbert space dimension of direct diagonalization scales unfavorably.

The investigation of systems with several atoms and many excitations is motivated by intriguing phenomena that already arise within the single-excitation subspace and the rotating-wave approximation.
App.~\ref{app:OBS-2GAs} and Fig.~\ref{fig:energy_density_two_giant_atoms} present two examples of this:
The former derives the formation of an oscillating bound state between two giant atoms.
The latter shows the emission of radiation from two giant atoms initially prepared as Bell states.
Depending on the relative phase of the Bell state, either all energy is radiated away into the waveguide or part of it remains bound in the field between the two atoms.
Future research directions include the investigation of superradiance, chiral quantum acoustics with and without an intrinsic time delay, and explicitly time-dependent models (see also \cite{duGiantAtomsTimedependent2022}), to implement gates between giant atoms.

At sufficiently strong couplings, it also becomes important to go beyond the two-level approximation and consider higher-lying excited states of the emitter.
Numerical simulations based on the matrix-product state ansatz can treat few-level emitters, and thus the techniques employed in this work can readily be adapted for future studies of non-Markovian dynamics beyond the two-level approximation at strong couplings.

\section*{Acknowledgements}
The authors would like to thank C. Rusconi for useful discussions.
R.H.J. gratefully acknowledges support by the Wenner-Gren Foundations.
J.K. gratefully acknowledges support from the European Union’s Horizon 2020 FET-Open project SuperQuLAN (899354) and the Deutsche Forschungsgemeinschaft (DFG, German Research Foundation) under Germany’s Excellence Strategy – EXC-2111 – 390814868.

\appendix

\section{Oscillating bound states of two giant atoms \label{app:OBS-2GAs}}

\begin{figure*}
\centering
	\includegraphics[width=.9\textwidth]{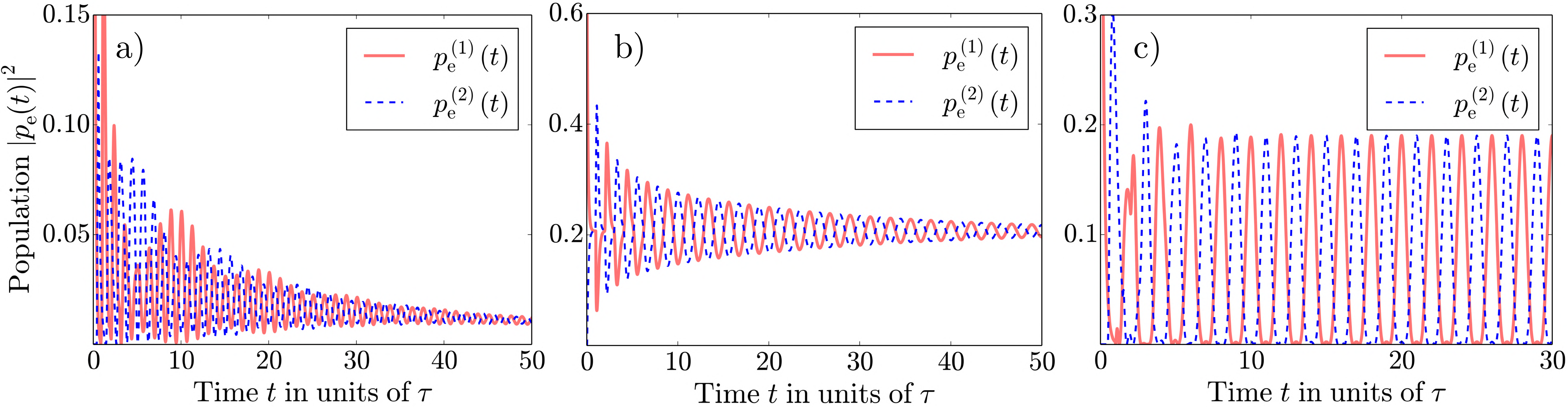}
	\caption{Population dynamics of coupled system.
	\textit{Red solid curves:} population dynamics $p_\mathrm{e}^{(1)}(t)$ of atom $1$, initially prepared in the excited state.
    \textit{Blue dashed curves:} population dynamics $p_\mathrm{e}^{(2)}(t)$ of atom $2$, initially prepared in the ground state.
    \textit{Parameters:} (a) $\Omega = 40\pi, \gamma\tau = 4, \ts/\tau = 0.39$; (b) $\Omega = 40\pi, \gamma\tau = 4, \ts/\tau = 0.025$; (c) $\Omega = 10\pi, \gamma\tau = \pi, \ts/\tau = 0.5$.}\label{fig:Decat2GA}
\end{figure*}

\begin{table}[b!]
\begin{center}
\begin{tabular}{|l || l | l|}
\hline
Case & Condition 1 & Condition 2 \\
\hline
\hline
Symmetric & $\frac{n\tau}{4\ts} \in \mathbb{Z}$ & $\Omega \tau = \frac{\pi n}{2} \frac{\tau}{\ts} - \gamma \tau \sin \left ( \pi n/2 \right )$ \\
\hline
Anti-symmetric & $\frac{n\tau}{4\ts} = k + \frac{1}{2}, k \in \mathbb{Z}$ & $\Omega \tau = \frac{\pi n}{2} \frac{\tau}{\ts} + \gamma \tau \sin \left ( \pi n/2 \right )$ \\
\hline
\end{tabular}
\end{center}
\caption{Conditions for the existence of symmetric and anti-symmetric dark states.}\label{tab:dark-conditions}
\end{table}

In this appendix, we study the dynamics of two giant atoms coupled to a common acoustic waveguide within the rotating-wave approximation and in the non-Markovian regime, and sketch the emergence of oscillating bound states in such setups.
Each atom is modelled as a two-level system and couples to the acoustic field at two points separated by $\tau$, see~Fig.\ref{fig:setup}.
For the purpose of this appendix, the total Hamiltonian of the system reads $H = H_0 + H_\mathrm{int}$, where~\citep{Guo2017}
\begin{equation}\label{eq:H0}
	H_0 = \sum_{j=1,2} \Omega \spl_j\smi_j + \sum_{\nu=r,l}\int\!\!\text{d}\w_p\, \w_p\, \aop_\nu(\w_p)\adop_\nu(\w_p)
\end{equation}
is the total energy of the quantum emitters and of the acoustic modes, and
\begin{eqnarray}\label{eq:Hint}
	H_\mathrm{int} &=& \sqrt{\frac{\gamma}{4\pi}} \sum_{\nu=r,l} \int\!\! \text{d}\w_p \Big  [ \smi_1 \adop_\nu(\w_p) e^{-\im k_\nu x_1}\left( 1+e^{-\im k_\nu \tau} \right ) \nonumber \\ \newline
    &+& \smi_2\adop_\nu(\w_p)e^{-\im k_\nu x_2}\left ( 1+e^{-\im k_\nu \tau} \right ) +\hc \Big ]
\end{eqnarray}
is the interaction between the emitters and the modes, with $\sigma^+ = |\mathrm{e}\rangle\langle\mathrm{g}|$, $\sigma^- = |\mathrm{g}\rangle\langle\mathrm{e}|$ and the relaxation rate $\gamma$.
Here the indices $r$ and $l$ refer to right-moving and left-moving modes, respectively.
In Eqs.~\eqref{eq:H0} and \eqref{eq:Hint}, $\Omega$ denotes the emitters' frequency, $k_r \equiv \w_p/c$ ($k_l \equiv -\w_p/c$) denotes the wavevector of the right (left) propagating mode, $c$ is the speed of sound, and $x_1 = 0$ ($x_2 = \ts$) is the position of the left-most contact point of the first (second) emitter.

We focus on the single-excitation subspace and are interested in parameter regimes which give rise not only to purely dissipative dynamics, but display additional features.
Taking into account the mirror symmetry of the setup, we make the ansatz
\bea\label{eq:dark-ansatz}
	\ket{\psi_\pm} &=& \beta^\pm (\spl_1\pm\spl_2) \ket{00}\ket{\text{vac}} \\
    & & + \int_\mathbb{R}\!\!\! \text{d}\w_p \alpha_\pm(\w_p)\spare{a^\dagger_r (\w_p)\pm a^\dagger_l(\w_p)} \ket{00}\ket{\text{vac}},\nonumber 
\eea
where $\beta^+$ and $\beta^-$ are associated with symmetric and anti-symmetric dark-state solutions, respectively.
By substituting this ansatz \eqref{eq:dark-ansatz} into the Schr\"odinger equation of the system,
integrating out the phonons and then applying a Laplace transformation to the resulting equation of motion, we obtain the probability amplitude
of the dark-state solutions we are looking for.
This procedure is a generalization of the result derived in Ref.~\cite{Guo2020} to two emitters.
In this way the results shown in Fig.~\ref{fig:Decat2GA} and discussed below were obtained, \ie from solving the coupled, time-dependent differential equations for the populations of two atoms and their time derivatives [$p_e^{(1)}$, $\dot p_e^{(1)}$, $p_e^{(2)}$ and $\dot p_e^{(2)}$] numerically.

As in the main text, we denote the separation between two legs of the same atom as $\tau$, while the position of the first leg of the second atom is located at $\ts < \tau$.
This geometry is also referred to as the braided configuration~\cite{Kockum2018}.
We find a set of criteria to judge whether dark states are present in the system.
These conditions are summarized in Table \ref{tab:dark-conditions};
a symmetric (anti-symmetric) dark-state solution exists if the corresponding conditions are fulfilled for any $n \in \mathbb{N}$.
In that case, the probability amplitudes of the symmetric and anti-symmetric solutions, respectively, will have the form
\bea
	\beta^+(t) & = & \frac{e^{\im \pi n t/2\ts}}{2}\Big\{1-\gamma\tau \Big[1+\pare{1+\frac{\ts}{\tau}}\cos(\pi n /2)\nonumber\\
	& &+ 2\im \frac{\ts}{\tau} \sin(\pi n /2)\Big]\Big\}^{-1},\label{eq:dark-solutions-1}\\
	\beta^-(t) & = & \frac{1}{2}\frac{e^{\im \pi n t/2\ts}}{1+\gamma \tau\spare{1-(1-\ts/\tau)\cos(\pi n /2)}}.\label{eq:dark-solutions-2}
\eea

An oscillating bound state in the two-atom setup can be found in cases where the dark-state conditions in Table \ref{tab:dark-conditions} are fulfilled for various $n$.
In Fig.~\ref{fig:Decat2GA}, we show the resulting dynamics in different parameter regimes, but all in the non-Markovian regime where $\gamma \tau > 1$.
While Fig.~\ref{fig:Decat2GA}(a) displays a fast decay of the initial excitation,
Figs.~\ref{fig:Decat2GA}(b) and Fig.~\ref{fig:Decat2GA}(c) show the emergence of dark states.
In the long-time limit, these do not decay despite their dissipative environment.
The setup corresponding to Fig.~\ref{fig:Decat2GA}(b) hosts a symmetric dark state for $n = 2$ (compare Table \ref{tab:dark-conditions}).
In Fig.~\ref{fig:Decat2GA}(c), one symmetric ($n = 10$) and two anti-symmetric ($n = 9,11$) dark states are present.
More explicitly, the long-time limit of the initially excited atom is given by $p_\mathrm{e}^{(1)}(t) = |\beta_{n=2}^+(t)|^2$ in Fig.~\ref{fig:Decat2GA}(b), and by $p_\mathrm{e}^{(1)}(t) =|\beta_{n=9}^-(t) + \beta_{n=11}^-(t) + \beta_{n=10}^+(t)|^2$ in Fig.~\ref{fig:Decat2GA}(c).

\section{Smearing functions for alternative coupling point profiles}\label{app:coupling_point_profiles}
As mentioned in Sec.~\ref{sec:setup}, other smearing functions than  the Gaussian profile~\eqref{eq:GaussianProfile} could be used to model the coupling points and  to capture the coupling's frequency dependence.
Some generic examples, all of which are normalized as $\integral{x}{}{}\pf(x)=1$, are
\begin{align}
	&\text{Lorentzian: } &&f_L(x) = \frac{d}{\pi(d^2+x^2)}, \nonumber \\
	&\text{Rectangle: } &&f_R(x) = \frac1{2d}\chi_{[-d, d]}(x), & \nonumber \\
	&\text{Dirac delta: } &&f_D(x) = \delta(x),
\end{align}
where $2d$ for the Lorentzian and the rectangle function represents the (effective) diameter of the coupling point.
In the calculation of the coupling coefficients $f_j$, assuming that $d \ll L$ we may replace  $\integral{x}{0}{L}$ in~\eqref{eq:fj_defn} by $\integral{x}{-\infty}{\infty}$ , and obtain
\begin{align}
    f_j^L &= -\ii \sqrt{\frac{|k_j|}{2L}} \ee{-|k_j| d}, \nonumber\\
    f_j^R &= -\ii \sqrt{\frac{|k_j|}{2L}} \frac{\sin(k_j d)}{k_j d}, \nonumber \\
    f_j^D &= -\ii \sqrt{\frac{|k_j|}{2L}}.
\end{align}
Here, the UV-divergence of the $\delta$-coupling becomes evident, since the coefficients $f_j^D$ diverge as $j\to\infty$, which makes it difficult to introduce a UV cutoff and limit the calculations to a finite number $N<\infty$ of field modes.

\section{Lanczos algorithm and its implementation}
\label{app:lanczos_math}

Here we review the Lanczos algorithm~\cite{paige_computational_1972,simon_analysis_1984,simon_lanczos_1984,parlett_symmetric_1998,saad_numerical_2011,Qiao2004,Qiao2005} which we used to cast the field Hamiltonian into the chain form~\eqref{eq:chain-mapped-field}.
The arithmetically exact form of the algorithm is severely impacted by roundoff errors in any numerical implementation~\cite{paige_computational_1972,simon_analysis_1984}, and hence reorthogonalization of the calculated vectors is necessary. Here, partial reorthogonalization~\cite{simon_lanczos_1984} provides a method to save the numerical costs of reorthogonalization by monitoring the loss in orthogonality over the iterative steps of the Lanczos algorithm and only triggering reorthogonalization where necessary. (Our numerical implementation of this method, as outlined in the following, is similar to~\cite{Qiao2004,Qiao2005} which, however, consider complex symmetric matrices $A$.)

In its simplest form, the Lanczos algorithm takes a Hermitian matrix $A$ and a starting vector $\vec{v}$ as inputs, and it returns two matrices $T$ and $Q$, such that $T$ is tridiagonal and $Q$ is unitary with
\begin{align}
    Q^\dagger A Q = T.
\end{align}
The columns of $Q = (\textbf{v}_1, ..., \textbf{v}_{n})$ correspond to the orthonormal basis vectors of the transformation, and $\vec{v}_1=\vec{v}/\|\vec v\|$.

The simple form of the algorithm is easily derived by noting that the $j$th column of the equation
$
	A Q = Q T
$
yields
\begin{align}
	A \vec{v}_j = \beta_{j-1}\vec{v}_{j-1} + \alpha_j \vec{v}_j + \beta_{j} \vec{v}_{j+1}\,.
	\label{eq:iteration_single_vector}
\end{align}
The version of the exact simple Lanczos algorithm which is most stable in numerical implementations is:
\begin{align}
	&\text{Lanczos algorithm:}\nonumber\\
	&\vec{v}_0 = 0; \beta_0 = 0;\nonumber\\
	&\vec{v}_1 = \vec{v} / \left\lVert \vec{v} \right\rVert;\nonumber\\
	&\text{for } j = 1 \text{ to } n:\nonumber\\
	&	\qquad \vec{w} = A\vec{v}_j- \beta_{j-1}\vec{v}_{j-1};\nonumber\\
	&	\qquad \alpha_j = \vec{w}^\dagger\vec{v}_j ;\nonumber\\
	&	\qquad \vec{r}_j = \vec{w} - \alpha_j \vec{v}_j ;\nonumber\\
	&	\qquad \beta_j =  \left\lVert \vec{r}_j \right\rVert;\nonumber\\
	&	\qquad \text{if } \beta_j = 0: \text{end};\nonumber\\
	&	\qquad \vec{v}_{j+1} = \vec{r}_j / \beta_j;\nonumber\\
	&\text{end}.\nonumber\\\nonumber
\end{align}
In practical implementations, the break condition can be replaced by $\beta_j<\epsilon$ for a sufficiently small bound.

\begin{figure*}
    \centering
    \includegraphics[width=.9\textwidth]{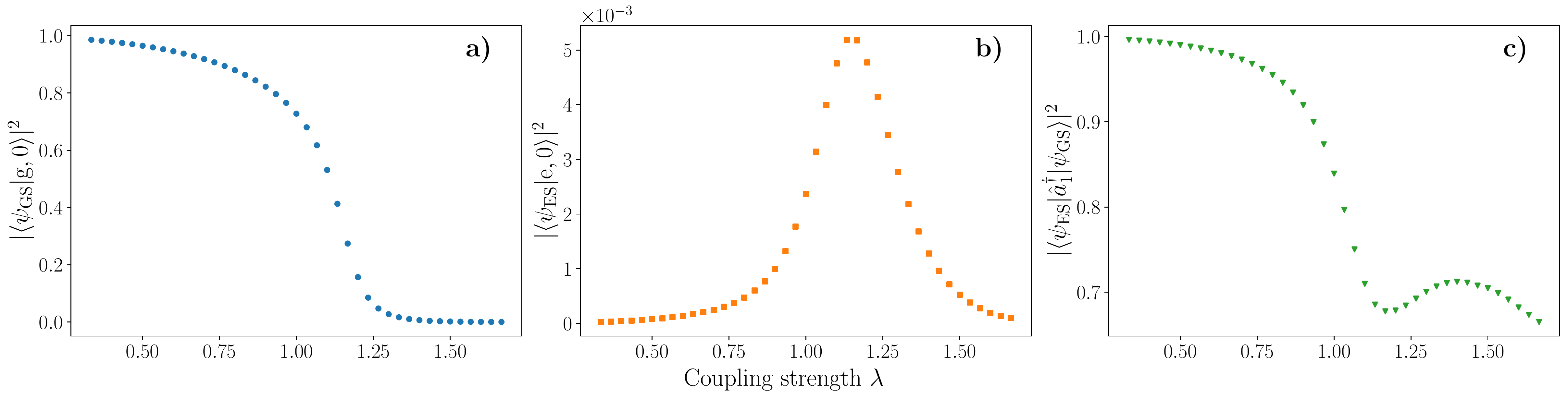}
    \caption{Overlap of  (a) $\ket{\psi_{\mathrm{GS}}}$ and  $\ket{\gup,0}$, (b)   $\ket{\psi_{\mathrm{ES}}}$ and $\ket{\eup,0}$, and (c) $\ket{\psi_{\mathrm{ES}}}$ and  $\hat a^\dagger_1\ket{\psi_{\mathrm{GS}}}$, as a function of the coupling strength $\lambda$.
    }
    \label{fig:overlap}
\end{figure*}

In finite precision arithmetic, rounding errors occur in~\eqref{eq:iteration_single_vector} which can be represented by an error vector,
\begin{align}
	A \vec{v}_j = \beta_{j-1}\vec{v}_{j-1} + \alpha_j \vec{v}_j + \beta_{j} \vec{v}_{j+1}+ \vec{f}_j\,.
	\label{eq:iteration_numerical_error}
\end{align}
Thus, defining $\xi_{k,j}=\vec{v}^\dagger_k\vec v_j$ as a symbol for the inner products of the iteratively obtained vectors, these no longer fulfill the ideal Kronecker relation $\xi_{k,j}=\delta_{k,j}$.
A key point is now that for the Lanczos algorithm to remain stable, it is not necessary to reorthogonalize all vectors, but it is sufficient to keep the $\vec v_j$ semi-orthogonal, \ie $\max_{1\leq k\leq j-1} |\xi_{k,j}|\leq\sqrt{\epsilon}$, for the roundoff unit $\epsilon$.
Hence, reorthogonalization is only required when this bound is violated at any iteration step of the algorithm.

The growth of the $\xi_{k,j}$ elements is determined by the recurrence relations~\cite{simon_lanczos_1984}
\begin{align}\label{eq:detect_orthogonality}
	\beta_j\xi_{k, j+1} &= \beta_k \xi_{j,k+1} + \alpha_k \xi_{j,k} - \alpha_j \xi_{k, j} \\ \nonumber &+ \beta_{k-1}\xi_{j,k-1}r- \beta_{j-1}\xi_{k, j-1} + \vec{v}_j^\dagger\vec{f}_k - \vec{v}_k^\dagger\vec{f}_j
\end{align}
together with $\xi_{j,j}=1$ and $\xi_{k,k-1}=\vec{v}_k^\dagger\vec{v}_{k-1}$. These, however, cannot be exactly calculated in numerical implementations since the error vectors $\vec{f}_k$ are not known.
Instead, the idea of partial reorthogonalization is to give an estimate for the terms 
$\theta_{k, j} \equiv\vec{v}_j^\dagger\vec{f}_k - \vec{v}_k^\dagger\vec{f}_j$ and $\xi_{j,j+1}$ by simulating them with random numbers,
\begin{align}
	\xi_{j, j+1} = n \epsilon \frac{\beta_1}{\beta_j}\Psi, \quad \Psi\in N(0, 0.6), \label{eq:adjacent_vectors_ortho}\\
	 \textbf{v}_j^\dagger\textbf{f}_k - \textbf{v}_k^\dagger\textbf{f}_j = \epsilon (\beta_k + \beta_j) \Theta,\quad  \Theta \in N(0, 0.3),
\end{align}
where $N(0, \chi)$ is a zero mean normal distribution with  variance $\chi$.
These estimates are then used in the original version of the algorithm to determine which vectors, if any, should be reorthogonalized at any given step of the algorithm~\cite{simon_lanczos_1984}. After a reorthogonalization has occurred, the relevant $\xi_{k,j}$ elements are reset to a normal distribution,
\begin{align}
	\xi_{k, j+1} = \epsilon \Xi, \quad \Xi  \in N(0, 1.5).
	\label{eq:reset_ortho}
\end{align}
For our purpose, we found the following simplified version to be sufficient, applying full orthogonalization to all vectors (we also used  wider normal distributions, as in~\cite{Qiao2004}):
\begingroup
\allowdisplaybreaks
\begin{align}
	&\text{Lanczos algorithm with partial orthogonalisation:}\nonumber\\
	&\textbf{v}_0 = 0; \beta_0 = 0;\nonumber\\
	&\textbf{v}_1 = \textbf{v} / \left\lVert \textbf{v} \right\rVert;\nonumber\\
	&\text{for } j = 1 \text{ to } n:\nonumber\\
	&	\qquad \textbf{w} = A\textbf{v}_j;\nonumber\\
	&	\qquad \alpha_j = \textbf{v}_j^\dagger \textbf{w};\nonumber\\
	&	\qquad \textbf{r}_j = \textbf{w} - \alpha_j \textbf{v}_j - \beta_{j-1}\textbf{v}_{j-1};\nonumber\\
	&	\qquad \beta_j =  \left\lVert \textbf{r}_j \right\rVert;\nonumber\\
	&	\qquad \text{Compute } \xi_{k, j+1} \text{ for } k = 1, \dots, j-1 \text{ using Eq. \eqref{eq:detect_orthogonality}};\nonumber\\
	&	\qquad \text{Set } \xi_{j, j+1} \text{ using Eq. \eqref{eq:adjacent_vectors_ortho}};\nonumber\\
	&	\qquad \text{Set } \xi_{j+1, j+1} = 1;\nonumber\\
	&	\qquad \text{if max}_{1\leq k \leq j}(|\xi_{k, j+1}|) \geq \sqrt{\epsilon}:\nonumber\\
	&	\qquad \qquad \text{Orthogonalise }\textbf{r}_j \text{ against } \textbf{v}_1, \dots, \textbf{v}_j;\nonumber\\
	&	\qquad \qquad \text{Perform orthogonalisation in the next iteration};\nonumber\\
	&	\qquad \qquad \text{Reset } \xi_{k, j+1} \text{ using Eq. \eqref{eq:reset_ortho}};\nonumber\\
	&	\qquad \qquad \text{Recalculate } \beta_j = \left\lVert \textbf{r}_j \right\rVert;\nonumber\\
	&	\qquad \text{if } \beta_j = 0: \text{end};\nonumber\\
	&	\qquad \textbf{v}_{j+1} = \textbf{r}_j / \beta_j;\nonumber\\*
	&\text{end}.%
\end{align}
\endgroup

For setups where~$b$ different emitters couple to the field, block Lanczos algorithms can be used to transform the field Hamiltonian.
The block Lanczos procedure takes a Hermitian matrix $A \in \mathbb{C}^{n\times n}$ and an orthonormal set of complex vectors $Q_1 = (\textbf{v}_1, \dots, \textbf{v}_b)$ as inputs. 
The algorithm then iteratively computes a unitary basis $Q = (Q_1, \dots, Q_p)$ and a block tridiagonal matrix $T$ such that 
\begin{align}
	Q^\dagger A Q = T  
	=\begin{pmatrix}M_1 & B_1^\dagger & 0 & \dots \\ B_1 & M_2& B_2^\dagger & \ddots \\0& B_2 & \ddots & \ddots \\ \vdots & \ddots & \ddots & \end{pmatrix},
\end{align}
where $M_i,B_i\in \mathbb{C}^{b\times b}$. The $M_i = M_i^\dagger$ are Hermitian, and the $B_i$ are upper triangular.
Analogously to the single vector Lanczos algorithm, we get the following procedure:
\begin{align}
	&\text{Block Lanczos algorithm:}\nonumber\\
	&p = n / b;\nonumber \\
	& Q_0, B_0 = 0;\nonumber \\
	&\text{for } j = 1 \text{ to } p:\nonumber\\
	&	\qquad Y = AQ_j;\nonumber\\
	&	\qquad M_j = Q_j^\dagger Y;\nonumber\\
	&	\qquad R_j = Y - Q_j M_j - Q_{j-1} B_{j-1}^\dagger;\nonumber\\
	&	\qquad \text{if } \text{max}(\lVert R_j\rVert) = 0: \text{end};\nonumber\\
	&	\qquad Q_{j+1} B_j = R_j; (\text{QR factorisation of } R_j)\nonumber\\
	&\text{end}.\nonumber\\\nonumber
\end{align} 
Also the block Lanczos algorithm needs to be stabilized in numerical implementations (\eg see~\cite{Qiao2005}).

\section{Overlaps of ground and first excited states}\label{app:overlap}

As described in Sec.~\ref{sec:stationary}, we numerically obtained the ground state of the coupled system $\ket{\psi_{\mathrm{GS}}}$ and its first excited state $\ket{\psi_{\mathrm{ES}}}$ for coupling strengths up to~$\lambda\lesssim1.8$. In addition to the discussion there, Fig.~\ref{fig:overlap} further characterizes these states by presenting their overlap with the states $\ket{\gup,0}$ and $\ket{\eup,0}$, respectively, as well as the overlap of the state $\hat a^\dagger_1\ket{\psi_{\mathrm{GS}}}$, obtained by applying the creation operator of the lowest energy eigenmode of the free field to  $\ket{\psi_{\mathrm{ES}}}$.

\section{Energy Density calculation}
\label{app:energy_density}

Figure~\ref{fig:energy_density_two_giant_atoms} shows the field energy density in the waveguide for a setup with two giant atoms. 
The underlying calculations and expressions are detailed in the following.
The energy density of the massless field in one dimension (1D),
\begin{align}\label{eq:edensity}
    \hat T_{00}(x)
    =\frac12\left(\hat \pi^2(x) +\left(\partial_x\hat\phi(x)\right)^2 \right)
    =\hat\pi_R^2(x)+\hat \pi_L^2(x)
\end{align}
is given by the sum of the left-moving and right-moving energy density, which in turn are given by the squares of the left- and right-moving sectors of the field momentum
\begin{align}
    \hat\pi_R(x) &=  \sum_{j\geq1}(-\ii)\sqrt{\frac{|k_j|}{2L}} \left( \ee{\ii\frac{2\pi j}L x}\hat a_j -  \ee{-\ii\frac{2\pi j}L x}\hat a_j^\dagger \right) \nonumber
    \\
    \quad \hat\pi_L(x) &=  \sum_{j\leq -1}(-\ii)\sqrt{\frac{|k_j|}{2L}} \left( \ee{\ii\frac{2\pi j}L x}\hat a_{j} -  \ee{-\ii\frac{2\pi j}L x}\hat a_{j}^\dagger \right).
\end{align}
Plugging this into~\eqref{eq:edensity} readily allows for the evaluation of the energy density expectation value from the covariance matrix of the field modes.

In the setup of Fig.~\ref{fig:energy_density_two_giant_atoms}, the initial states of the two atoms are entangled Bell states and the initial state of the field is the vacuum, \ie the system starts in the product state $\ket{\Psi_\pm}\otimes\ket0$. In Fig.~\ref{fig:energy_density_two_giant_atoms}(a), the atoms are initialized in the triplet state $\ket{\Psi_+}$, whereas in Fig.~\ref{fig:energy_density_two_giant_atoms}(b), they are initialized in the singlet state $\ket{\Psi_-}$,
\begin{align}
    \ket{\Psi_{\pm}}=\frac1{\sqrt2}\left(\ket{\eup_1\gup_2}\pm\ket{\gup_1\eup_2}\right).
\end{align}
The calculations for Fig.~\ref{fig:energy_density_two_giant_atoms} were performed within the RWA.  Hence the time evolution is restricted to the single-excitation subspace and,  in the evaluation of the expectation value of $\hat T_{00}(x)$, terms can be discarded that do not conserve the excitation number, \ie only terms of the form $\hat a_i^\dagger \hat a_j$  need to be taken into account.

The plots of Fig.~\ref{fig:energy_density_two_giant_atoms} show that for the singlet state $\ket{\Psi_-}$ all energy quickly radiates away from the atoms in the waveguide. However, when the atoms are initialized in the triplet state $\ket{\Psi_+}$ then the interference between the atoms' braided coupling points results in a significant amount of energy remaining bound between the outer pairs of  coupling points.

\bibliography{thesis}
\end{document}